\documentstyle[aps,epsfig]{revtex} 

\begin{document} 
 
\draft 
\title{A dual point description of mesoscopic superconductors} 

\author{E.~Akkermans$^1$, D.M.~Gangardt$^1$  and K.~Mallick$^2$ 
 \\$^1$Department of Physics, 
 Technion, 32000 Haifa, Israel \\  $^2$Service de Physique Th\'eorique, 
 CEA Saclay, 91191 Gif-Sur-Yvette, France} 
\maketitle 

\begin{abstract} 
  We present  an   analysis of the magnetic response 
 of a mesoscopic superconductor, {\it i.e.} a  system 
  of sizes comparable to the coherence length and to  the London 
 penetration depth. Our approach  is based  on  special 
 properties of  the two dimensional Ginzburg-Landau equations, 
 satisfied at the dual point $(\kappa = \frac{1}{\sqrt{2}}).$ 
 Closed expressions for the free energy and the magnetization 
 of the superconductor are derived. A  perturbative analysis in the vicinity 
 of the dual point allows us to  take into account vortex interactions, 
 using a new scaling result for the free energy. 
 In order to  characterize the 
 vortex/current interactions, we 
 study vortex  configurations that are  out 
 of thermodynamical equilibrium. Our predictions 
 agree with the results of recent experiments performed 
 on mesoscopic aluminium disks. 
\end{abstract} 


\section {Introduction}

 The ability  to detect and manipulate vortices with great 
 sensivity in  systems  of small size such as mesoscopic superconductors 
  \cite{geim1} or atomic condensates \cite{dalibard} 
  has generated an outgrow of 
 interest in   the mechanism of creation and annihilation 
 of vortices and in the  study  of stable and metastable 
  vortex  configurations. In particular, 
 recent advances in the technique of Hall magnetometry \cite{geim0} 
  have allowed to measure the 
  magnetization of small superconducting samples containing 
 only a few vortices  \cite{geim1,geim2}. These   experiments 
   are conducted on aluminium disks  well below the superconducting 
 transition temperature, whereas   previous measurements were 
 performed only  in the vicinity of 
   the normal/superconductor phase boundary \cite{moschalk,buisson}. 
    Besides, the magnetization measurements in \cite{geim1} 
  are carried out  on an 
 individual disk   and  not on an ensemble of disks as in \cite{buisson}. 
 The  radius $R$  and the thickness $d$  of the sample  used in the experiments 
  are comparable to   the   superconducting 
    characteristic lengths, 
  {\it i.e.} the London penetration 
 length ($\lambda =$ 70nm) and the coherence length ($\xi = $250nm). 
 Such a  sample can neither  be considered 
 to be  macroscopic, nor microscopic. The system falls, rather,  in 
   a  {\it mesoscopic} regime  where   surface effects 
 are of the same order of magnitude as the bulk effects. Thus, the 
 magnetic response of a mesoscopic 
   superconducting disk   to an applied   field 
  depends strongly  on its size and is very different from that 
 of a macroscopic superconductor. 
 When  the radius $R$ 
 of the sample is much smaller  than the coherence length $\xi$, 
 no vortex can nucleate, 
 the  normal/superconductor phase transition 
 is second order and the 
  magnetization $M$,   as a function of the external 
 applied field $H_e$,   has a  non-linear  behaviour 
 (non-linear Meissner effect \cite{peetreview,deoprl1}). 
 If   $R$  is  comparable 
 to $\xi$, the superconducting phase transition 
 is first order and a bistable hysteresis  region 
 appears in the $M-H_e$ curve. 
 For    $R$  greater than  $\xi$, 
 the phase  transition 
 is again second order, and when the applied field exceeds 
 a critical value   $H_{1}$, 
 the magnetization curve exhibits 
 a  series of discontinuous jumps 
  corresponding  to  the successive entry of vortices into  the sample. 
  This qualitative interpretation is supported, at least for low  applied 
 magnetic fields, by the periodicity   of the jumps which corresponds 
 to the entrance of an additional superconducting quantum  of flux 
into the disk. 
 For larger fields, or equivalently for higher density of vortices, 
 both the period  and the height  of the jumps become smaller, 
 a behaviour related  to the  interactions between 
 the vortices and to transitions between stable vortex configurations, 
 with the same  number of vortices. 
 
   The magnetization shows also a  hysteretic behaviour depending 
 on the direction of the field sweep, due to the presence of a  confining 
 energy barrier (the absence of remanent  magnetization precludes 
 pinning effects). In  some metastable states, the sample may 
 exhibit even  a paramagnetic response \cite{geim2} 
 whereas in thermodynamic equilibrium 
 a superconductor is diamagnetic. 
 
  These experimental results have led to a  renewed interest in 
  the theory of mesoscopic superconductors. 
 Numerical computations have shown that 
 the phenomenological  Ginzburg-Landau  theory is well suited to describe 
   a superconducting sample in the mesoscopic regime, 
 even  far from the critical temperature. These works have 
  revealed   physical phenomena that 
  play an important role in such systems 
 (for a  review see \cite{peetreview}), like: 
 the role of surface barriers for vortex nucleation and hysteresis 
 \cite{deo,pala1,pala2}; 
 the interplay between vortex/vortex   and 
 vortex/edge interactions  that explains   vortex 
 structures  in mesoscopic disks \cite{pala1,deo}; 
 the transition between a giant multiple vortex state 
 and a state with several vortices carrying a unit quantum of flux 
 \cite{schweig}.

    The Ginzburg-Landau free energy of a superconductor 
 involves two fields, the  (complex) order parameter 
 $ \psi = |\psi| e^{i\chi}$ and the vector potential 
  ${\vec A}$. The minimization of this free energy 
 leads to  a set of  two coupled non-linear partial differential 
 equations for $\psi$   and ${\vec A}$, involving 
  the two characteristic 
 lengths  $\lambda$ and $\xi.$ 
But the solutions depend only on one relevant  number, 
 the phenomenological  Ginzburg-Landau parameter $\kappa$ defined by
\nopagebreak 
  \begin{equation} 
 \kappa = \frac{\lambda}{\xi} \; . 
 \end{equation} 
  A  macroscopic superconductor is said to be of type~I 
   if  $  \kappa <  \frac{1}{\sqrt{2}} $  and of   type~II 
 if  $  \kappa >  \frac{1}{\sqrt{2}}.  $  A macroscopic  
    superconductor of Type II  admits  a stable 
  Abrikosov vortex lattice phase 
  when the applied field  $H_e$ lies between 
 the first penetration  field  and the upper critical field 
 \cite{degennes}. 
   For  aluminium,  $\kappa$ is smaller than 
  $\frac{1}{\sqrt{2}}$,  hence a macroscopic 
  sample of {\it Al} is a type~I superconductor. 
 
 Analytical studies of the  Ginzburg-Landau equations in two-dimensional systems require 
 the use of  various approximations since, in general, exact solutions can not 
 be found due to  the non-linearity. 
 One approach   is to linearize  the equations  assuming  
 $ |\psi| \ll 1, $ and   to decouple  them 
   by supposing that  the magnetic field $ B$ 
 in  the sample is equal to the applied field 
 $H_e$. This approach describes correctly   the superconducting/normal 
 phase  boundary 
  \cite{moschalk,buisson,little,groff,zwerger}, but fails to 
  explain the behaviour of the sample deep inside 
 the superconducting state. For example,  in  the linearized theory, 
 all  the vortices are  at the center of the disk 
 \cite{zwerger}  and therefore one can not study 
 the role of surface barriers, 
 the interaction between vortices, and  the fragmentation of a giant vortex 
 into unit  vortices.  Besides, the  critical  fields corresponding to 
  the successive  entrance of vortices into 
 the sample do not scale correctly 
  with the size of the system 
 ({\it e.g.}  experimentally, 
 the  entrance field $H_{1}$ of the first vortex  scales 
 as ${R}^{-1}$ whereas  the  linear theory 
 predicts a  ${R}^{-2}$ dependence). Of course, 
 in the vicinity of the upper critical field  \cite{zwerger} 
 the linearized theory 
 agrees quantitatively with the experimental results. 
 
 A second  approach is to use the 
   London  equation  which can be  derived  from the Ginzburg-Landau 
 equations  by supposing 
 that $ |\psi| = 1$  everywhere except on 
 a finite number of  isolated points, called 
 vortices, where  $ |\psi| = 0.$ 
  London's equation is valid 
 rigorously   when the parameter $\kappa$ goes to infinity, 
 {\it i.e.} for extreme type II superconductors 
 in which vortices are indeed point-like. 
 Many theoretical results have been  derived from  the  London 
 equation, such as 
  discrete  nucleation   of flux lines in a  thin 
 cylinder \cite{bobel,shapoval}  or in a thin disk  \cite{pearl,fetter}, 
 the existence of  surface energy barriers \cite{degennes,bean},  and 
  the computation of  polygonal ring configurations of vortices 
  in finite samples 
 \cite{russes,bresiliens}. However,  when  $\kappa \to \infty,$ 
 the minimum energy is obtained for one flux quantum 
 per vortex \cite{sarma,bogo} and 
   vortices have a  hard-core  repulsive interaction impeding 
 the formation of a giant vortex state. 
 Moreover, the surface  energy barriers calculated from 
 London's equation 
 are quantitatively different from those obtained 
 by numerically  solving the Ginzburg-Landau equations \cite{peet2}. In fact, 
 the experimental 
 conditions  are    far off  the London limit, 
 although  thin {\it Al} disks are likely to  have an effective 
 $\kappa$ greater  than its measured  value \cite{geim1}  of  0.28 
 (in a thin disk, one can argue, following \cite{pearl}, 
 that the effective  London length is of the order of  ${\lambda^2}/{d}$, 
  and this results 
 in a higher value of $\kappa$).

 We  follow another 
 approach,  less explored in the literature, based on an exact 
 result for the  two-dimensional  Ginzburg-Landau  equations. 
  In an infinite plane 
 reduce to first order differential equations that can be decoupled 
  when the parameter $\kappa$ takes the special value $ 1/{\sqrt{2}},$ 
 called the dual point\cite{sarma,bogo,harden}. At that  point, 
 the free energy is a topological invariant 
 of the system. In  \cite{akmal}, 
 we  generalized  this method 
 to a finite domain with boundaries; this enabled us to classify 
 solutions with  different number of vortices and to derive analytical 
 expressions  for the free energy and the magnetization of a mesoscopic 
 disk as a function of the applied field. 
 Our results agreed qualitatively  with the experimental data, 
 and  even quantitatively 
 when the number of vortices in the system is low. 
 However, some important features such as the non-linear Meissner 
 effect in a fractional fluxoid disk,  the variation 
 of the amplitude and the period of the jumps  in 
 the $M-H_e$ curve 
 could not be described.  Moreover, in  \cite{akmal}, 
 we  discussed only  the case where  $R$ is much larger 
 than $\xi$ and did not obtain the different regimes 
 of the magnetization curve when the ratio $R/\xi$ is varied.

 In this paper, we study 
 the Ginzburg-Landau free energy ${\cal F}$ not only 
 at the dual point   $\kappa = 1/{\sqrt{2}}$ but also   in  its  vicinity 
 where  vortices start to  interact weakly \cite{rebbi}. 
 Taking into account non-linear effects, our calculations 
 describe  the magnetic response of the sample 
 as  its size changes, providing an 
 understanding of the  non-linear 
 Meissner effect and of the multi-vortex state. We  shall also study 
 non-equilibrium vortex configuration in order to determine  
 the interaction between a vortex and edge currents. 
 
  The plan of this paper  goes as follows: in section 2, some basic 
 features of the Ginzburg-Landau theory of superconductivity are 
 recalled. In section 3, after studying  the case of an infinite 
 system, we generalize the Bogomol'nyi's approach  to a finite size 
 superconductor  and  calculate its  free energy 
 at the dual point.  This result is applied to 
 an infinite cylinder in section 4. The case of a mesoscopic disk is studied 
 in section 5 and magnetization curves are obtained for systems 
 of different sizes. In section 6, we obtain the free energy and the 
 magnetization of  a cylindrically symmetric system 
  when $\kappa$ is close to  the dual point. 
  The  surface energy barrier  for a  one  vortex  state out of thermodynamic 
 equilibrium is calculated in section 7. 
 In the last section we discuss our results and suggest some 
 further generalizations. Some mathematical details are included 
 in the two appendices.

 \section{The  Ginzburg-Landau theory  of  superconductivity} 
 
  We  recall here  some basic features of the 
  Ginzburg-Landau theory  and   define our notations. 
  The order parameter  $\psi = |\psi| e^{i\chi}$ is a complex number  
 and the potential vector ${\vec A}$ satisfies  
 ${\vec{\nabla}}\times{\vec A} = {\vec B}$, where ${\vec B}$ is 
 the local magnetic induction. 
 The two characteristic lengths $\lambda$ and $\xi$ 
 appear as phenomenological parameters. 
  In this work,  we   measure  lengths in units of 
 $\lambda { \sqrt 2} $, the magnetic field 
 in units of $\frac{\phi_0}{4\pi\lambda^2}$ and the vector potential 
 in units of  $\frac{\phi_0}{2\sqrt{2}\pi\lambda}$ where  
the flux quantum $\phi_0$ is given by 
  $\phi_0 = \frac{hc}{2e} .$ 
  The  Ginzburg-Landau free  energy $\cal F$,  defined  as  
 the difference of the free energies ${\cal F} = F_S (B) - F_S (0)$,    
is measured in   units of $\frac{H_c^2}{4\pi}$ 
 where ${H_c}$    the thermodynamic field satisfies  
${H_c} = {\sqrt 2} \kappa \frac{\phi_0}{4\pi\lambda^2}$. 
 In these units,  $\cal F$  is given by   
\begin{equation} 
{\cal F} =  {\int_{\Omega}} {1 \over 2} |B{|^2} + {\kappa^2} 
 |1 - |\psi{|^2}{|^2} + 
|({\vec{\nabla}} - i{\vec A})\psi {|^2}   \,\,\,\,\,\, , 
\label{energ} 
\end{equation} 
where the integration is over the superconducting domain $\Omega$. 
 The  Ginzburg-Landau equations   that minimize  ${\cal F}$,  become 
\begin{eqnarray} 
  -  ({\vec{\nabla}} - i{\vec A})^2 \psi & = &  2{\kappa}^2 \psi 
(1 - |\psi{|^2})   \label{adgl1} \\ 
    {\vec \nabla} \times {\vec B} & = & 2 {\vec \jmath} 
  \label{adgl2} 
\end{eqnarray} 
 Equation (\ref{adgl2})  is the   Maxwell-Amp\`ere 
 equation   with  a  current density 
${\vec \jmath}= \mbox{Im}( {{\psi^*}} {\vec \nabla} \psi ) - |\psi{|^2}{\vec A}$  
related to the  superfluid velocity ${\vec v_s}$ by  
\begin{equation} 
{\vec v_s} = {{\vec \jmath} \over |\psi{|^2} } = {\vec{\nabla}} \chi - {\vec A} 
\label{vsuper} 
\end{equation} 
    Outside the superconducting sample, 
  $\psi = 0.$  The boundary condition 
 on the surface of the superconductor   is obtained by requiring 
 that the normal component of the current density vanishes 
 (superconductor/insulator boundary condition  \cite{degennes}): 
\begin{equation} 
({\vec{\nabla}} - i{\vec A})\psi |_{\hat{\bf n}} = 0 
\label{bondcond} 
\end{equation} 
 here ${\hat {\bf n}}$ is the unit vector 
 normal at each point to  the surface 
 of the superconductor. 
 
  The London  fluxoid is  
  the quantity 
$ \left(  \displaystyle{ {\vec \jmath} \over { |\psi|^2} } 
 + {\vec A}   \right), $ 
    that  is identical to  ${\vec \nabla}\chi$. 
   Since  $\chi$ is  the phase of the  univalued function 
 $\psi ,$ the  circulation of   the London  fluxoid along  a closed contour 
 ${\cal C}$  is quantized  \cite{degennes,london}: 
 \begin{equation} 
 \oint_{\cal C} ( \frac {{\vec \jmath}}{ |\psi{|^2}} + {\vec A}).{\vec dl} 
 =     \oint_{\cal C} 
 {\vec \nabla}\chi .{\vec dl} =  2\pi  n 
\label{quantize} 
 \end{equation} 
 The integer $n$ is the winding 
 number of the phase of the system  along the contour ${\cal C}$ 
 and is a topological characteristic of the system.

 In this study, the superconducting sample 
 is either an infinite cylinder or a thin disk, 
 with cross-section of radius $R$,   placed in an external magnetic 
 field parallel  to its axis. Since $R$ is an important 
  parameter, we define the  dimensionless quantity: 
\begin{equation} 
    a = \frac{\lambda\sqrt{2}}{R} 
 \label{pmtra} 
\end{equation} 
  $a$ is  supposed to be small compared to 1 (typically $a \sim 1/10$ 
 in the experiments) 
  unless stated otherwise. 
 The flux created by the external and uniform magnetic 
   field $H_e$ (expressed  in units of 
 $\frac{\phi_0}{4\pi\lambda^2}$) through the cross section $\pi R^2$ 
 of the sample is equal to 
 $ \pi R^2 H_e \frac{\phi_0}{4\pi\lambda^2} = \frac{H_e}{ 2 a^2} \phi_0 .$ 
 The  flux  $\phi_e$,  in  units  of the flux quantum $\phi_0$, 
 is thus given by: 
  \begin{equation} 
   \phi_e =   \frac{H_e}{ 2 a^2} 
  \end{equation} 
 We emphasize that, in the  units we have chosen, 
 the flux $\phi_b$ of a magnetic field 
  $\vec B$ through a surface $\Omega$ is obtained via the following 
 formula: 
 \begin{equation} 
 \phi_b = \frac{1}{2\pi} \int_{\Omega} {\vec B}.{\vec dS} = \frac{1}{2\pi} 
 \oint_{\partial\Omega} {\vec A}.{\vec dl} 
 \label{defflux} 
\end{equation} 
  An extra  factor $1/{2\pi}$ appears here because $B$ is given 
 in units of  $\frac{\phi_0}{4\pi\lambda^2}$, the surface 
 in units of $2\lambda^2$ and  the flux  in units of $\phi_0$. 
 
  Since  we are studying a superconductor  in an applied external field, 
   the relevant thermodynamic potential is the Gibbs 
 free energy  $ G$ obtained  from  $ F$ 
 via a Legendre transformation: 
 \begin{equation} 
  G =   F  -  H_e \int_{\Omega} B =   F 
  -  H_e  2\pi \phi_b =  F  - 4\pi a^2 \phi_e \phi_b 
   \label{gibbsener} 
\end{equation} 
 In a normal sample,  $\psi = 0$ and $B = H_e$. Therefore, 
  the Gibbs free energy $ G_N$ of a normal sample 
 is  given by: 
\begin{equation} 
  G_N  =  F_N -  H_e \int_{\Omega}B  = F_N -2 \pi a^2 \phi_e^2 
 \label{gibbsnormal} 
\end{equation} 
    At  thermodynamic equilibrium, 
 the superconductor  selects the  state of minimal  Gibbs free energy. 
  The  quantity that we are interested in, and which is measured 
 in experiments, 
  is the magnetization  
   $M$ of the superconductor due to  the applied field given by $4 \pi M =  
B - H_e $.  
It is obtained, at thermodynamic equilibrium and  
up to a constant equal to the superconducting condensation energy, 
 from the difference of the (dimensionless) Gibbs energies  
\begin{equation} 
{\cal G} = G_S - G_N = {\cal F} + 2 \pi a^2 \phi_e^2 - 4\pi a^2 \phi_e \phi_b 
\label{gsgn} 
\end{equation}  
using  the 
thermodynamic relation \cite{degennes}: 
\begin{equation} 
 -  M = \frac{1}{2\pi} 
 \frac{\partial {\cal G}}{\partial\phi_e} 
 \label{magthermo} 
\end{equation}

 \section{Free energy of a superconductor at the dual point}

 We now study the particular case of the dual point, 
 defined by   $\kappa = {1 \over \sqrt 2} .$ 
  For this value of the Ginzburg-Landau parameter, 
  the free energy (\ref{energ}) 
 of a two dimensional domain  ${\Omega}$  can be written as 
 \cite{bogo,akmal}: 
\begin{equation} 
{\cal F} = {\int_{\Omega}} \left( {1 \over 2} \left( B - 1 + |\psi|^2\right)^2 + 
  |{\cal D} \psi {|^2} \right)  
 + {\oint_{\partial \Omega}} ({\vec\jmath} + {\vec A}).{\vec dl} 
 \label{identitebog} 
\end{equation} 
 where  the   operator $\cal D$ is 
defined as  ${\cal D} = \partial_x + i\partial_y -i(A_x + iA_y)$ and 
 the second integral   is over the boundary 
 of the domain ${\Omega}.$ 
 
\subsection{The case of an infinite system} 
 
  If we suppose that the domain ${\Omega}$ is infinite 
 and superconducting at 
  large distances \cite{bogo}, 
 {\it i.e.} $ |\psi| \to 1$ at infinity,  then 
 the boundary integral in  (\ref{identitebog}) 
 is identical to   the fluxoid. Using 
  the quantization property 
 (\ref{quantize}), we  obtain 
\begin{equation} 
{\cal F} = 2 \pi n +   {\int_{\Omega}} \left({1 \over 2} (B - 1 + |\psi|^2)^2 + 
  |{\cal D} \psi|^2 \right) 
\end{equation} 
 The free energy is thus minimum  when Bogomol'nyi equations 
 \cite{bogo} are satisfied, that is when, 
\begin{eqnarray} 
 {\cal D} \psi &=&  0 
 \label{equabogo1} \\ 
    B  &=&  1 - |\psi|^2 
\label{equabogo2} 
\end{eqnarray} 
  Thus,  the total free energy    results only 
 from the boundary term in (\ref{identitebog}) 
 and is a purely topological number: 
\begin{equation} 
{\cal F} = 2 \pi n \,\,\,\, . 
\end{equation} 
 The free energy is   proportional to the number of vortices: 
 at the dual point, vortices do not interact with each other 
 \cite{bogo,rebbi}. 
 
\subsection{Finite size systems} 
\label{sectgamma}
In a finite system with boundaries, 
  vortices do not interact with each other 
 at the dual point but  they  are repelled 
 by the edge currents. Therefore, at  thermodynamic equilibrium, 
 all vortices collapse into a giant vortex state. 
 Since the  superconductor under discussion has a circular cross-section, 
 this giant vortex (or {\it multi-vortex}) is located at the center 
 and the system is invariant under cylindrical symmetry. 
  In  a {\it finite}  size mesoscopic 
 superconductor at the dual point, 
  the   boundary integral, 
    in (\ref{identitebog}),  can not be identified 
  with  the fluxoid  because 
 $|\psi|$ is in general  different from 1 on the boundary of the system. 
  This quantity 
  is no more  a topological  integer  but 
 a continuously varying real number. 
 The two terms of  (\ref{identitebog}) can  not, therefore, 
 be minimized {\it separately} 
 to obtain the optimal  free energy. 
  In \cite{akmal}, we found a method to circumvent this difficulty: 
 if  the system is invariant under cylindrical symmetry, {\it i.e.} 
  all the vortices are at the center of the disk,  then 
  the current  density has only an azimuthal component 
 $\jmath_{\theta}$. The current  $\jmath_{\theta}$ 
  has opposite signs near the center (where the vortex is located) 
 and at the edge  of the disk (where   Meissner currents 
 oppose  the penetration of the external field). 
 Hence,   there exists a circle $\Gamma$ on which 
 $\jmath_{\theta}$   vanishes \cite{akmal}. Along  $\Gamma$, 
 we have 
\begin{equation} 
   {\vec\jmath} + {\vec A} =   {  {\vec \jmath}  \over  {|\psi|^2} } 
 +  {\vec A} =  {\vec \nabla}\chi \,\,\,\, \hbox{ and therefore } 
 \,\,\,\, {\oint_{\Gamma}} ({\vec\jmath} + {\vec A}).{\vec dl} 
 = 2\pi n 
\label{ptegam} 
\end{equation} 
 The domain  $\Omega$  can thus be divided  into  two sub-regions, 
 $\Omega~=~{\Omega_1}\cup {\Omega_2},$ 
such  that the boundary between  ${\Omega_1}$ 
 and ${\Omega_2}$  is the circle   $\Gamma$. 
 By convention, we 
 call ${\Omega_1}$ the bulk and 
 the annular ring ${\Omega_2}$  the boundary region (see 
 figure 1).

  Numerical solutions of the Ginzburg-Landau equations in a two 
 dimensional superconductor, with cylindrical symmetry, 
 clearly show the  separation of the sample cross-section 
 into two distinct subdomains. 
  In figure 2, we have plotted the order 
 parameter and the magnetic field in a cylinder of radius 
 $R =  10 \lambda\sqrt 2 $ with one vortex at the center. These two 
 quantities vary only near the center and near  the edge: there is a whole 
 intermediate region in which $|\psi|$  and $B$  
remain almost constant. When the system is large enough, these constant values 
 are,  to an excellent precision, identical to the asymptotic values 
 of $|\psi|$  and   $B$  in an infinite system.  
 The current vanishes for  a  value of 
 $r$ for which  $\frac{dB}{dr} = 0$, and this 
 determines  the  radius  of the circle  $\Gamma.$ 
 In figure 2, this 
 corresponds  approximately to $ r\simeq 5.5$, though practically, the circle 
 $\Gamma$  can be placed anywhere in the saturation region where 
 the current is infinitesimally  small.

Although the Bogomol'nyi equations (\ref{equabogo1}, \ref{equabogo2}) do not minimize the  
Ginzburg-Landau free energy \cite{dunne}, we notice that  
the behaviour of  $|\psi|$  and $B$ in the bulk 
 subdomain  ${\Omega_1}$  is still given by the relation  
$ B  =   1 - |\psi|^2 , $(\ref{equabogo2}) which, as shown  
in figure 3, represents indeed an excellent approximation.  
 Thus,  using  (\ref{identitebog}) 
 and (\ref{ptegam}), we conclude that  at the dual point 
 the free  energy   of ${\Omega_1}$ can be calculated as 
   that of 
    an infinite domain, namely 
\begin{equation} 
  {\cal F}({\Omega_1})  = 2 \pi n 
\label{free1} 
\end{equation} 
 We also emphasize that the flux in ${\Omega_1}$   is  quantized 
 and one has: 
\begin{equation} 
 \frac{1}{2\pi} \int_{\Omega_1} B = n 
\label{flux1} 
\end{equation}

 To calculate $ {\cal F}(\Omega_2)$, the identity (\ref{identitebog}) 
 valid at the dual point is of no use anymore, 
 since  $|\psi|$ is in general different from 1  at  the boundary, 
 and the boundary integral in (\ref{identitebog}) can not 
 be identified to the fluxoid. We therefore have to go back 
 to the definition (\ref{energ}) of the  Ginzburg-Landau free energy 
 which becomes at the dual point: 
\begin{equation} 
  {\cal F}(\Omega_2)= \int_{\Omega_2} {  B^2 \over 2} 
  + (\nabla |\psi|)^2 
 + |\psi|^2  |{\vec{\nabla}}\chi - {\vec A}|^2 
  + { (1 - |\psi|^2)^2  \over 2}
\label{exp2} 
\end{equation} 
The assumption  of  cylindrical symmetry implies 
 that $\chi = n\theta$ where $\theta$ is the polar angle 
 and   $n$ the number of  vortices  present  at the center of the disk. 
  Examining again figure 2, we observe that in  $\Omega_2$, 
 the order parameter and the magnetic field vary  from their 
  values on the edge  to their saturation values over a region of  width 
 $\delta$, which is of order 1  
 in units of $\lambda\sqrt{2}$\footnote{Indeed,   
 one has for a thick system  
 $\delta \simeq \lambda$  at the dual point.  For  a thin  film  
of  thickness $d,$  
 $\delta \simeq \lambda^2/d$   in the London limit \cite{pearl}. 
  Since  we are considering a 
  mesoscopic regime in which    $d \simeq   \lambda$, both expressions 
  indicate that   $\delta$  is of order   1.}  
 The  length   $\delta$ therefore  
 represents the typical distance  over which the 
 integrand  in   (\ref{exp2}) has a non negligible value. 
 
 With the help of this observation, we shall  estimate $ {\cal F}(\Omega_2)$ 
   using  a variational  Ansatz: 
 we shall consider that  the modulus of the order parameter has a constant 
 value  $\psi_0$ over  a ring of width  $\delta$, 
 included  in $\Omega_2 $ and  that ${\vec A}$ and  
 ${\vec B}$  decay exponentially with  a characteristic length $\delta$ 
 from their  
 boundary value to their bulk value. 
 Clearly,  our approximation  will be 
   valid  only if  the width of   $\Omega_2$ 
 is  large enough compared  to 1. We first remark that our Ansatz 
 is  compatible with the boundary condition (\ref{bondcond}), 
  which reduces  here to $\frac{d\psi}{dR}= 0$ and that it allows us  
 to neglect the curvature term $(\nabla |\psi|)^2$ in (\ref{exp2}). 
 To evaluate the term proportional 
 to the superfluid velocity  
$v_s (r)$ (\ref{vsuper}), we first notice that, 
 due to the Meissner effect, it decreases from the  
boundary at $r = R$ with a behaviour well described by  
\begin{equation} 
v_s (r) = v_s (R) e^{- (R - r)/ \delta}  
\end{equation} 
with $v_s (R) = a (n - \phi_b)$. To obtain the last equality we used that the  
 boundary value of the vector  potential  is 
${\vec A}(R) =  a  \phi_b  {\hat u_{\theta}}  ,$ 
where $\phi_b$  is the total flux through the system. 
Hence,  for a constant amplitude $\psi_0$ of the order parameter, we have 
\begin{equation} 
   \frac{1}{ 2 \pi}  {\cal F}(\Omega_2) =  
   \frac{\delta}{2a} \left(\psi_0^2 v_s^2 (R) 
 +( 1 - \psi_0^2)^2 \right) \,\,\,\, + \frac{1}{ 2 \pi} 
  \int_{\Omega_2} {  B^2 \over 2} 
\label{avecb} 
\end{equation} 
 
The magnetic contribution in (\ref{exp2}) is obtained from the  
typical  magnitude $\bar{B}$ of the magnetic field in  $\Omega_2$ determined 
 using  (\ref{defflux})  and (\ref{flux1}) as 
\begin{equation} 
  \phi_b = \frac{1}{2\pi} \int_{\Omega} B = \frac{1}{2\pi} \int_{\Omega_1} B 
 + \frac{1}{2\pi} \int_{\Omega_2} B = n + \frac{\delta}{a}\bar{B} 
\end{equation} 
Thus, using  the fact that 
 $B^2 $ decrease exponentially with a characteristic length $\delta/2$, 
  we  estimate the contribution of the magnetic energy 
 to ${\cal F}(\Omega_2)$ as being: 
 \begin{equation} 
\frac{1}{ 2 \pi}  \int_{\Omega_2} {  B^2 \over 2} =  \frac{\delta}{2a} 
 \frac{ \bar{B}^2}{2} = \frac{a}{4\delta}  (n - \phi_b)^2 
\end{equation} 
 After  substituting  this expression into (\ref{avecb})  we  minimize 
   $ {\cal F}(\Omega_2)$ with respect to $\psi_0$. 
    The optimal variational  value of  $\psi_0$ is given by: 
\begin{equation} 
 \psi_0^2 = \left\{\begin{array}{lc}
1 - {1 \over 2} v_s^2 (R) = 1 - {a^2 \over 2} (n - \phi_b)^2 & 
\hbox{     if  }\;\;
 |a(n - \phi_b)| \le {\sqrt 2} \\ 
0    & \hbox{     if  }\;\;
 |a(n - \phi_b)| > {\sqrt 2}
\end{array}\right.
\label{meilphi} 
\end{equation} 
 Inserting  these expressions in (\ref{avecb}), we obtain   the  variational  
 free energy ${\cal F}(\Omega_2)$: 
 \begin{equation} 
  \frac{1}{ 2 \pi} {\cal F}(\Omega_2) = \left\{\begin{array}{lc}  
   A v_s^2 (R) - B v_s^4 (R) & 
\hbox{ if  }\;\;  |a(n - \phi_b)| \le {\sqrt 2}  \\ 
  \frac{\delta}{2a} 
     + \frac{1}{4a\delta}  v_s^2 (R) &  
  \hbox{ if  }\;\;  |a(n - \phi_b)| >  {\sqrt 2}
\end{array}\right. 
 \label{freebord2} 
  \end{equation} 
with $A$ and $B$ defined by 
\begin{eqnarray} 
A &=& {\delta \over 2 a} \left( 1 + {1 \over 2 \delta^2}\right) \nonumber \\ 
B &=& {\delta \over 8 a} 
\label{AB} 
\end{eqnarray} 
 
The total free energy of 
 the mesoscopic superconductor containing $n$ vortices, 
 at the dual point, is thus: 
 \begin{equation} 
  \frac{1}{ 2 \pi} {\cal F}(n ,\phi_b) = n+\left\{\begin{array}{lc}  
   A v_s^2 (R) - B v_s^4 (R) & 
\hbox{ if  }\;\;  |a(n - \phi_b)| \le {\sqrt 2}  \\ 
  \delta/2a 
     +   v_s^2 (R)/4a\delta &  
  \hbox{ if  }\;\;  |a(n - \phi_b)| >  {\sqrt 2}
\end{array}\right. 
 \label{enlibre} 
  \end{equation}

   This energy is the sum of two contributions: 
 
  (i) a {\it bulk}  term  proportional to $n$ which is a  topological 
  quantity at the dual point. 
 
  (ii) A  {\it boundary } term, reminiscent  of  the well-known 
 `Little and Parks' free energy \cite{degennes} 
 (this  boundary term  can  be given a geometric 
 interpretation in terms of  a geodesic curvature  \cite{akmal,houches}).

\section{Free energy and magnetization of a cylinder at the dual point} 
 
 We   now  apply the relations (\ref{enlibre}) to the simple case of 
  an infinitely long superconducting  cylinder  of  radius $R > \lambda$, 
 lying in  an  external  field $H_e$  directed along its axis.   There 
 are two contributions to the total flux $\phi_b$: 
 the flux of  $n$ vortices 
 present at the center of the sample 
  and  a  fraction of the applied flux $ \phi_e$ localized 
 near the boundary and proportional to $\lambda/R$ 
 (due to the Meissner effect). Hence, 
 \begin{equation} 
  \phi_b = n + 2a \phi_e  \,\,\,  \hbox{ with } \,\,\, 
  \phi_e = \frac{  H_e}{2 a^2} 
\label{fluxcyl} 
\end{equation} 
   The exact numerical coefficient in front of the term $a \phi_e$ 
 does  not affect the result of  our calculation; we take  it equal to 2, 
  the value obtained in   the London limit \cite{bobel}. 
  The total free energy, 
 using (\ref{enlibre}) and the fact that  $v_s(R)=-2a^2\phi_e$,   
is  given by 
 \begin{equation} 
  \frac{1}{ 2 \pi} {\cal F}(n ,\phi_b) = n+\left\{\begin{array}{lc}  
  4a^4 \left(A \phi_e^2 - 4a^4 B \phi_e^4 \right) & 
\hbox{ if  }\;\;  a^2|\phi_e| \le {1/\sqrt 2}  \\ 
  \frac{\delta}{2a} 
     + \frac{a^3}{\delta} \phi_e^2 &  
  \hbox{ if  }\;\;  |a^2|\phi_b| >  {1/\sqrt 2}
\end{array}\right. 
 \label{energcyl} 
  \end{equation}
Using (\ref{gibbsener}) and (\ref{energcyl}), 
 the Gibbs free energy, $ {\cal G}(n, \phi_e ) ,$   of 
  a cylinder containing  $n$ vortices at the dual point  is given by: 
\begin{equation} 
\frac{1}{ 2 \pi} {\cal G}(n, \phi_e )  = n \left( 1 - 2 a^2 \phi_e \right) 
  + P(\phi_e) 
\label{gibcyl} 
\end{equation} 
  where $ P( \phi_e )$ is a polynomial in $ \phi_e  $ 
 that  does not depend on $n$.  Hence, 
 all the curves ${\cal G}(n, \phi_e )$ 
  meet  at 
\begin{equation} 
  \phi_c = \frac{1}{2 a^2} 
\end{equation} 
  For values of $ \phi_e $ less than this  critical value, 
 the free energy  is minimized  if there are  no vortices. 
  At $ \phi_e = \frac{1}{2 a^2} $ all  vortices are nucleated simultaneously 
  and the sample  becomes normal. 
 This    value corresponds to a 
 critical  applied field $H_e$ which is  equal to 
 1 in our units, or   restoring the  units back, 
 and recalling  that $\kappa = 1/\sqrt{2}$ 
 \begin{equation} 
  H_e = \frac{\phi_0}{4\pi\lambda^2} =\frac{\phi_0}{2\sqrt{2}\pi\lambda\xi} 
\end{equation} 
 This is precisely the formula 
 for  the  thermodynamic critical field of a superconductor \cite{degennes} 
 (which, for a cylindrical superconductor with $\kappa \le \frac{1}{\sqrt{2}},$ 
  is the same as  the upper critical field). 
  The magnetization $M$ of the cylinder satisfies  the  linear 
 Meissner effect: 
\begin{equation} 
 -  M= \frac{1}{2\pi} 
{  {\partial {\cal G}(n, \phi_e ) }  \over  {\partial \phi_e}  } 
 = H_e( 1 - c a ) \,\,\,\, \hbox{ with }  \,\,\,\, 
 c = 4  -\delta\left(1 + \frac{1}{2\delta^2}\right) \,\,\,. 
\end{equation} 
  The  macroscopic result \cite{degennes} is $ -M = H_e$; the 
  finite-size correction to the susceptibility is 
 proportional  to $R^{-1}$. 
 
 Thus, the  well-known  results for   an infinite superconducting 
 cylinder can easily be retrieved  from the   dual point approach. 
  We  now  proceed to the  study  of the magnetic response of a thin disk.

\section{A mesoscopic  disk at the dual point} 
 
 To modelize the experimental sample  of \cite{geim1,geim2}, we 
  consider a mesoscopic disk of thickness $d$ 
 smaller  than $\xi$ and $\lambda$.  Because the disk is very thin,  we take 
 the order parameter and the magnetic field  to be   constant 
 across the thickness $d$ of the sample \cite{akmal}.  This enables 
 us to study the disk as an effective two-dimensional system. 
 However, unlike the case of a long cylindrical sample, 
  strong  demagnetization effects 
 are present in  a thin disk. 
  The value of $B$ near the edge  of the disk is larger 
  than the applied field $H_e$ because 
  geometric demagnetization  effects  induce  a 
 distortion  of the flux lines \cite{deo}. 
 Hence the continuity  condition $B(R) = H_e$  (\ref{fluxcyl}) 
  valid for a long cylinder    does not apply to describe a thin disk. 
   
In order to find a more suitable choice for the boundary condition for a thin disk,  
we notice that the higher value of the magnetic field at the boundary, a feature which has 
 been  obtained  from   numerical computations  \cite{schweig2}, results from a  
 demagnetization factor $\cal N$ close to one, such that \cite{degennes}  
$H = {H_e \over 1 - {\cal N}} $ 
in the Meissner phase. The  flux lines are distorted by 
 the sample and they pile up near the edge of the disk. To describe this,  
we shall thus take as  
 boundary condition 
 for a  thin disk, the expression proposed in  \cite{akmal}, which consists in 
  taking the potential-vector 
 at the edge of the disk equal to its  applied value, 
 {\it i.e.} 
\begin{equation} 
  {\vec A}(R) = {\phi_e} a {\hat u_{\theta}} 
\end{equation} 
or 
\begin{equation} 
   \phi_b = \phi_e  \label{bcdisk} 
\end{equation} 
 Again, this relation does not mean that the field $B$ is uniform 
 and equal to its external  strength. A more refined value for the boundary condition  
could have been obtained by  
using the expression ${\cal N} \simeq 1 - {\pi \over 2} {d \over R}$ in the limit  
$d \ll R$ of a flat disk. Then, $H \simeq {2 R \over \pi d} H_e$ or equivalently  
 $\phi_b \simeq {4 \delta \over d} \phi_e$. But, since $\delta \simeq d$, we shall use for convenience  
the simpler boundary condition given above.

 Substituting $v_s (R)=a(n-\phi_e)$ in (\ref{enlibre}),  the free 
 energy ${\cal F}(n , \phi_e) $  of a thin disk containing $n$ vortices 
  is found to be 
 \begin{equation} 
  \frac{1}{ 2 \pi} {\cal F}(n ,\phi_b) = n+\left\{\begin{array}{lc}  
 A a^2 (n-\phi_e)^2-B a^4 (n-\phi_e)^4 & 
\hbox{ if  }\;\;  a|(n-\phi_e)| \le {\sqrt 2}  \\ 
  \frac{\delta}{2a} 
     + \frac{a}{4 \delta} (n-\phi_e)^2 &  
  \hbox{ if  }\;\;  a|(n-\phi_e)| >  {\sqrt 2}
\end{array}\right. 
 \label{energdisk} 
  \end{equation}
 and the corresponding  Gibbs free energy is obtained using 
 (\ref{gsgn}). 
 In  our previous work \cite{akmal}, we obtained an expression which can be retrieved 
 from  (\ref{energdisk}) by taking  $\delta=1$ and by neglecting the  
 magnetic energy as well as  
 the $a^3$ term. Despite  these crude approximations, our 
 analytical results agreed satisfactorily with experimental data, though they 
  could not describe  neither the behaviour of 
 a disk with a radius smaller than $\lambda$ and $\xi$, 
 nor its behaviour when $R$ is increased. 
  We apply our present approach  to  a thin disk with a radius $R$ 
 much smaller than $\xi$, and then we consider the case  $ R > \xi .$ 
 
 \subsection{Fractional fluxoid disk and Non-Linear Meissner Effect} 
 
  We now consider a disk small enough so that no vortices can nucleate 
 {\it i.e.}  its radius $R$  is less than  $ \xi$ 
 (such a system is sometimes called 
  a {\it  fractional  fluxoid} disk \cite{peetreview}). 
 If there are  no vortices, 
 the domain $\Omega_1$ is  empty and  $\Omega =\Omega_2$. 
  Since the radius  of   $\Omega$ is small with respect to both 
 $\lambda$ and $\xi$, we can no more use 
  the expression (\ref{energdisk}) 
  for the free energy, but we can assume 
 that the amplitude $|\psi|$ of the order parameter has a uniform 
value $\psi_0$ all over the disk and that 
the magnetic field  equals the external applied 
field $B = H_e$. Moreover, in the  absence 
 of vortices  ${\vec \nabla} \chi =0$ and we can choose the Landau  
gauge $A(r) = rB/2$.  Starting 
  from  (\ref{exp2}), and after minimizing the free energy with respect to $\psi_0$, we find 
 the difference between the free energies 
   of the superconducting   and   the normal states  to be: 
\begin{eqnarray} 
  {{\cal G} \over 2 \pi } &=&  \frac{\phi_e^2}{4} 
 \left( 1  - \frac{a^2}{4} \phi_e^2  \right) 
  \,\,\, \hbox{ if  } \,\,\, 
   a \phi_e \le \sqrt{2}  \nonumber \\ 
 {{\cal G} \over 2 \pi } &=&  0  \,\,\, \hbox{ otherwise  } \,\,\, 
\end{eqnarray} 
 From (\ref{magthermo}) we deduce the magnetization $M$  of the sample: 
 \begin{eqnarray} 
 -  M &=&  \frac{1}{2\pi} \frac{\partial {\cal G}}{\partial\phi_e} = \frac{1}{2} 
 \left(  \phi_e - {a^2 \over 2} \phi_e^3 \right) 
  \,\,\, \hbox{ if  } \,\,\, 
   a \phi_e \le   \sqrt{2} \nonumber \\ 
 M &=&  0   \,\,\, \hbox{ otherwise  } \,\,\, 
\label{mag0} 
\end{eqnarray} 
  The curve representing this magnetization is a cubic. 
  The upper critical field is  
 $\phi_e = 1/a$, i.e. $H_e \propto R^{-1}$; 
  this scaling  agrees with  the 
 linear analysis of \cite{zwerger} in the limit 
 $ R \ll \xi$. 
  The transition between 
 the superconducting phase  and the normal phase is of second order. 
 In figure 4, we plot the relation (\ref{mag0})  for 
 $-M$ as a function of the external flux $\phi_e$. The dots represent 
 the experimental points obtained from \cite{geim2}. The analytical curve 
 has been scaled so that the maximum value of the magnetization 
  and the critical flux coincide with the corresponding experimental data.

 \subsection{Mesoscopic disk with vortices} 
 
 We now consider a disk with $R \ge \xi$. 
 The Gibbs free energy difference ${\cal G}(n, \phi_e)$  of the disk with $n$ vortices 
 is given by (\ref{gsgn}). 
 The entrance field $H_n$  of the  $n$-th vortex is 
 obtained  by solving the equation ${\cal G} (n ,\phi_e) ={\cal G}(n-1 , \phi_e)$ 
 which, using (\ref{energdisk}), reduces to 
 \begin{equation} 
  \frac{2}{\delta} = 
 a(1 + \frac{1}{2\delta^2})\left((n -1- \phi_e)^2 
- ( n - \phi_e )^2 \right)  - \frac{a^3}{4} 
  \left((n -1- \phi_e)^4 - ( n - \phi_e )^4 \right) 
 \label{chp1} 
 \end{equation} 
  Using  the following  change of variable 
 \begin{equation} 
 \phi_e = n -  \frac{1}{2} +  \frac{y}{2a} 
  \label{chgt} 
 \end{equation} 
  we obtain  an equation for $y$ 
 \begin{equation} 
      \frac{2}{\delta} =(1 + \frac{1}{2\delta^2}) y - \frac{y^3}{8} 
 \label{eqy} 
  \end{equation} 
 (a  term $a^2/8$  has been neglected in comparison  to 1). 
    The 
   solution   of  (\ref{eqy})  that  satisfies 
 $y \ge 0$ (because  $\phi_e \ge 0$)   depends on 
  the value of the parameter $\delta$. One can show that  the polynomial 
 $P(y) = (1 + \frac{1}{2\delta^2})y - \frac{y^3}{4} -   \frac{2}{\delta} $ 
 always has a positive root.  We  retain only the smaller 
 positive root $y_0$   of (\ref{eqy}) because in thermodynamic equilibrium, 
 the system always chooses the state with minimal Gibbs free energy. 
 Restoring the usual units, and using (\ref{chgt}), 
 the nucleation fields are found to be: 
\begin{eqnarray} 
 {H_1} &=&  y_0 { {\phi_0} \over {2 \pi {\sqrt 2} R \lambda} } 
 + { {\phi_0} \over {2 \pi  R^2} }  \nonumber \\ 
   H_{n +1} &=&  H_1 + n  \frac{\phi_0}{\pi R^2} 
\label{chpcritique} 
\end{eqnarray} 
   When the applied field $H_e$ lies between $H_n$ and 
 $H_{n+1}$, the disk contains exactly $n$ vortices 
  and its magnetization  is  calculated using (\ref{magthermo}). 
 In figure 5, we have plotted the magnetization of a mesoscopic disk 
 with $ R= 10 \lambda \sqrt 2$ both from exact numerical solutions of the Ginzburg-Landau equations  
and from the expression (\ref{magthermo}). The agreement is very satisfactory.   
For larger values of the number $n$ of vortices, a discrepancy between the theoretical  
and the numerical expressions appears which results from the interaction between the  
vortices and the edge currents that we have neglected until now. 
 
 The expression (\ref{enlibre}) is also in  good  agreement 
 with previous experimental and numerical results \cite{geim1,peetreview}. 
 A  non-linear Meissner behaviour still exists before the 
 nucleation of the first vortex as well as between successive jumps. 
 The field $H_1$ of nucleation of the first vortex scales as 
 $R^{-1}$. 
 The transition between a state with $n$ vortices to a state 
 with $(n+1)$ vortices is of first order  since the 
   entrance of a new vortex induces a jump in the magnetization. 
 These  jumps are  of constant height and 
 have  a  period $ \frac{\phi_0}{\pi R^2 }$. 
 If we use the experimental values of \cite{geim1} for $R$ and $\lambda$ 
 we obtain a value for the period of the jumps which is in very good agreement 
 with the experimental value.  
 
  If $R$ is smaller than a threshold value, 
 the system  is a  fractional fluxoid disk 
 with a second order phase transition. 
 If $R \simeq  1$, a vortex can nucleate in the disk and a first order 
 transition occurs. When $R$ increases, the number of jumps increases (as 
 $R^2$).  These  qualitative changes  of behaviour with increasing  $R$, which are the  
important features  
 obtained from the present model, 
 have been indeed observed in experiments carried out on disks of different sizes. 
 In an earlier study \cite{akmal}, we obtained satisfactory values for the 
 nucleation fields but the fractional fluxoid disk, and the different 
 regimes obtained by  increasing $R$ could not be explained because 
  we neglected 
 subdominant terms that are  retained  here.

  It has been observed experimentally  that 
 the period and the height  of the jumps cease to be  constant 
 when the number of vortices increases. 
These effects   are  related both to  interactions 
 between the vortices and between vortices and edge currents. The purpose of the next section 
 is to take into account these  interactions 
 and to obtain a better estimate for the free energy and the magnetization 
 of a mesoscopic disk.

 \section{Weakly interacting vortices in the vicinity of the dual point} 
 
   So far  we have obtained analytical expressions 
 for the free energy and the magnetization of a thin superconducting 
 disk {\it at the dual point.}  When the Ginzburg-Landau parameter 
 has the  special value,  $\kappa = \frac{1}{\sqrt{2}},$  vortices 
 do not interact. This fact, discussed  in 
 \cite{bogo,rebbi},  implies that the bulk free energy does 
 not depend on the location of the vortices. However, when 
  $\kappa$  is away from the dual point,  the vortices start interacting 
 among themselves;  therefore  the bulk free energy 
   ceases to be a purely topological 
  integer $n$ and the  vortex interaction  energy must 
  be taken into account. 
 Because of this interaction 
 the vortices are no longer necessarily   placed at the center 
 of the disk: in an equilibrium configuration, 
  the cylindrical symmetry can be broken and 
 the  optimal free energy  may correspond to geometrical  patterns such as 
 regular  polygons, polygons with a vortex at the  center, 
 or even rings of polygons \cite{pala1,bresiliens,peet2}. 
 It is   the  competition between the  interaction 
 amongst vortices and the interaction between vortices and edge currents 
 that  determines  the shape of the equilibrium configuration.

 Analytical studies  were mostly carried out 
  in the limit  $\kappa \to \infty$  and were 
 based on the London equation 
 \cite{bobel,fetter,bresiliens} for which  
  vortices  are point-like and  have a hard-core repulsion  \cite{degennes}. 
   We shall study a regime where    $\kappa$ is slightly 
 different  than $\frac{1}{\sqrt{2}}$, {\it i.e.} 
   a  regime where vortices 
 interact {\it weakly}. We shall  determine, to the leading order in 
 $(\kappa - \frac{1}{\sqrt{2}}),$ the interaction energy of the 
 vortices.

 \subsection{The interaction energy}

     In order  to obtain  an 
 estimate for the free  energy of a system of interacting vortices, 
  we have solved numerically the Ginzburg-Landau equations 
 for a cylindrically symmetric  infinite  system  with $n$ vortices 
 located at the center (these  equations  
 are explicitly written  in Appendix A). 
 The free energy per vortex  is plotted 
 in figure 6 as a function of $\kappa$, for 
 $n=1,2,3,5$ and 10. 
 At the dual point, the free  energy per vortex 
  is equal to 1 and is independent of $n$: all the curves pass through 
 this  point.   When 
  $\kappa$ is different from $\frac{1}{\sqrt{2}}$ the interaction 
 between the vortices changes the value of the free energy. One can deduce 
 from figure 6 that vortices attract each other for 
 $\kappa$ less than $\frac{1}{\sqrt{2}}$ while they repel each other when 
 $\kappa \ge \frac{1}{\sqrt{2}} .$

 From  our numerical results 
 we observed that in the vicinity of the dual point, 
  the free energy ${\cal F}(\kappa, n)$ 
   satisfies the following 
 scaling behaviour: 
 \begin{equation} 
 \frac{1}{2\pi} {\cal F}(\kappa, n) = n (\kappa\sqrt{2})^{\alpha(n)} 
 \label{fscaling} 
  \end{equation} 
 We note  that  the relation (\ref{fscaling}) is 
 exact at the dual point. 
 For $n=1$, ${\cal F}(\kappa, 1)$ is 
 nothing but the self energy $ {\cal U}_S $ 
 of a vortex. In the vicinity of the dual 
 point we  can write: 
 \begin{equation} 
 \frac{1}{2\pi} {\cal F}(\kappa, 1) =  1 + \alpha(1)(\kappa\sqrt{2} - 1) 
 \label{f1scaling} 
  \end{equation} 
  The values of the function $\alpha(n)$, as determined 
 from numerical computations, 
 for $n$ ranging from 1 to 30  are given in the table \ref{alpha}. 

 We can now  derive  an approximation 
 for the free energy of a $n$ vortices configuration  located 
 at the center 
 of the disk and  for  $\kappa$ close to $\frac{1}{\sqrt{2}}.$ 
 Since this configuration is cylindrically symmetric, one can again use 
  the circle $\Gamma$ to separate the system into two subdomains 
  $\Omega_1$ and 
  $\Omega_2$  and then estimate separately the 
 two contributions to the total free energy. 
 From  our  numerical scaling result, we deduce a  formula 
 for the bulk free energy of a finite system which is 
 valid for $\kappa$ close to the dual point. 
  Expanding  (\ref{fscaling}) 
 in the vicinity of the dual point, we obtain: 
  \begin{equation} 
  \frac{1}{2\pi} {\cal F}(\Omega_1) = n  + (\kappa\sqrt{2} - 1) n \alpha(n) 
 \label{scalingbulk} 
\end{equation} 
 and the boundary contribution, 
  obtained via a variational Ansatz is now given by: 
  \begin{equation} 
  \frac{1}{ 2 \pi}  {\cal F}(\Omega_2)  =\left\{\begin{array}{lc}
 A v_s^2 (R) -  B(\kappa) v_s^4 (R)  & \hbox{ if  } \;\;\;\; 
|a(n - \phi_e)| \le 2 \kappa   \\ 
\kappa^2\delta/ a  + v_s^2/4a\delta &
\hbox{ if  } \;\;\;\; |a(n - \phi_e)| > 2 \kappa \end{array}\right.
\label{pertbord} 
  \end{equation} 
where $A$ is still given by the relation (\ref{AB}) while $B(\kappa)$ is now given by $B(\kappa) = \delta /  
16 a \kappa^2$. 
 
 The magnetization curve of figure 7 shows both the numerical results and a plot of the  
magnetization deduced from  (\ref{pertbord}) using (\ref{magthermo}). 
 We notice that the magnetization of a mesoscopic disk 
 is modified when the interactions 
  between vortices are taken into account. The period 
 and the amplitude of the jumps are not constant anymore; 
 besides, the non-linearity of the curve  between two successive jumps 
 is enhanced. 
 These important features of the $M-H_e$  curve 
  were  observed in previous experimental 
 and numerical results \cite{geim1,deoprl1}. Here we have shown that 
 these features are a consequence of vortex interactions.

\subsection{Two-body interaction energy} 
 
The exponent $\alpha(n)$ in the relations (\ref{fscaling}) or (\ref{scalingbulk})  
allows to describe the interacting potential between vortices.   
It is interesting to compare the result (\ref{fscaling}) with the 
energy of $n$ vortices obtained by assuming a two-body interaction. 
 In this case the  energy of the whole system  
 of $n$ vortices can be written as a sum of two terms 
\begin{equation} 
\frac{1}{2\pi}{\cal F} = n  {\cal U}_S 
  + \frac{n(n-1)}{2} {\cal U}_I (0) 
\end{equation} 
 where  ${\cal U}_S$ represents, as noted before, the  self-energy of a vortex 
 and  ${\cal U}_I$ the two body interaction potential. 
 Using the data of \cite{rebbi}, we can estimate these 
 two energies to the leading order in $(\kappa\sqrt{2} -1)$. 
 We obtained: 
 \begin{eqnarray} 
 {\cal U}_S  &=& 1 + \beta_1 (\kappa\sqrt{2} -1) 
 \,\,\,\,\,  \hbox{ with }  \,\,\,\,\, \beta_1 \simeq 0.4 \,\, 
 \label{selfenerg}  \\ 
{\cal U}_I (r) &=&  \beta_2 (\kappa\sqrt{2} - 1) 
 \min\left( 1, \exp(-C(r - \frac{1}{\kappa})) 
 \right)    \,\,\,\,\,   
 \label{interac} 
\end{eqnarray} 
with $\beta_2 \simeq \frac{1}{4}$ and $ C  \simeq \frac{1}{2}$. 
 From this analysis, and assuming only two-body interaction, 
  we derive  an approximate value for the free energy 
of a configuration with $n$ vortices placed at the same point: 
  \begin{equation} 
\frac{1}{2\pi}{\cal F}(\kappa, n) = n  {\cal U}_S 
  + \frac{n(n-1)}{2} {\cal U}_I (0) \simeq 
 n  +     (\kappa\sqrt{2} -1)  n 
  \left(  \beta_1 + \beta_2 \frac{n-1}{2} \right) 
  \label{pertbulk} 
\end{equation} 
  If we compare this relation  to the previous expression 
 (\ref{scalingbulk}) we find that instead of   the sublinear function $\alpha(n)$ 
  we have a linear behaviour 
 $\beta_1 + \frac{n-1}{2}\beta_2$. 
  Hence, the  function $\alpha(n)$ takes into 
 account not only two-body interactions among vortices 
 but also multiple interactions which are present for values of $\kappa$ around the dual point  
unlike the large $\kappa$ limit where only the two-body contribution remains.

 \section{vortex/edge interactions in system without cylindrical symmetry}

   In this  section, we  calculate the energy  at the dual point 
    of a system   with only one vortex  that is not located 
  at the center of the disk. Such a configuration 
 is not in   thermodynamic equilibrium and its free energy can 
 be related to a  surface energy barrier 
 (analogous to the classical Bean-Livingston barrier in  the London limit). 
 We first show that even when the   cylindrical symmetry is broken, 
  the system can still be separated into bulk and  edge 
 domains.

 \subsection{Bulk and edge domains. The curve $\Gamma$} 
 
   We have seen in section \ref{sectgamma}   that 
 when  one or more    vortices are located 
 at the center of the disk,   there 
 exists  a circle $\Gamma$ on which the current  vanishes 
 identically. This circle  allowed us  to define  a bulk and an edge 
  domain and to identify the bulk energy with the fluxoid. 
 
  If  all the vortices are not placed   at the center of the disk 
 ({\it i.e.} the configuration is not cylindrically symmetric) 
  there is in general  no  curve   of zero 
 current. However the curve $\Gamma$ has now 
 the following property:  at   each point  $M$ of  $\Gamma$ 
   the  current ${\vec\jmath}$    is normal 
 to  $\Gamma$.  The existence 
 of such a curve is shown by  the following  argument. 
  Consider a  disk  with only 
 one vortex $V$  situated at a point different from the center 
 of the disk. Take a   line  segment joining 
 the vortex $V$ to the closest  point $S$  on  the  boundary of the disk 
 (see figure 8). 
 The component of the  current density 
 normal to the  $VS$  segment  changes its sign 
  when one goes from $V$  to 
  $S$. Hence,  there exists  a point $M$    along this segment 
  where  the current  either vanishes or  is parallel 
  to  $VS.$  To draw 
 the curve $\Gamma$ we  start   from $M$  in a direction  orthogonal to the 
 $VS$ segment, and then   $\Gamma$  is constructed via infinitesimal steps 
 by imposing that at a point $M' = M + dM$, very close to $M$,  the 
 direction of  $\Gamma$ is orthogonal to the direction of the current 
 at $M'.$ 
 
 Although we lack a general proof, we believe 
 on topological grounds  that  for vortices at arbitrary positions, 
 there always exists a $\Gamma$ curve which is everywhere 
  orthogonal to the current (one should note  that 
  $\Gamma$ does not necessarily have 
 only one connected component). In \cite{dima}, we shall present a numerical 
  construction of    $\Gamma.$  In the sequel of this work we assume 
 that $\Gamma$ exists, that it encircles all the vortices, and consists 
 of one or many  simple  closed curves. 
 We shall call the curve $\Gamma$  {\it the separatrix.} 
 
  Using $\Gamma,$ 
 the domain $\Omega$  can be decomposed   in 
two regions $\Omega_1$ and $\Omega_2$  such that: 
 
 (i) $\Omega_1\bigcup\Omega_2 =  \Omega$ ; 
 
 (ii)  $\Omega_1$ contains all the vortices ( $\Omega_1$ 
  may have multiply connected components); 
 
 (iiii)  $\Omega_2$ contains the edge  of the disk; 
 
  (iv) the separatrix $\Gamma$ is the boundary 
  between  $\Omega_1$ and $\Omega_2$  and is 
  everywhere normal to  the current density.

  The remarkable property of the separatrix  implies  that 
  along  $\Gamma$  one can write: 
\begin{equation} 
 \oint_{\Gamma} ({\vec\jmath} + {\vec A}).{\vec dl} = 
    \oint_{\Gamma} \left(      {  {\vec \jmath}  \over  {|\psi|^2} } 
 +  {\vec A} \right).{\vec dl}  =  \oint_{\Gamma} {\vec \nabla}\chi.{\vec dl} 
\label{separatrix} 
\end{equation} 
  since along  $\Gamma$, $  {\vec\jmath}.{\vec dl} = 0$. 
 Since the separatrix is the boundary of  $\Omega_1$, the 
   property (\ref{separatrix})  ensures  that   the total 
 magnetic flux through  $\Omega_1$ is quantized. 
  Hence, at the dual point, we  can again use the method of 
  Bogomoln'yi  and find  the free energy of $\Omega_1$ to be  a purely 
 topological number, just as for an infinite domain, even if the 
 cylindrical symmetry is broken.

  \subsection{Free energy of one vortex:  the  surface energy barrier.}

As before, we estimate the contribution 
  ${\cal F}(\Omega_2)$ 
 to the total free energy via a variational 
 Ansatz, taking the modulus of the order parameter to be  constant. 
To obtain a qualitative result for the surface energy barrier we neglect 
the magnetic energy so that, at the dual point, we have: 
\begin{eqnarray} 
\frac{1}{2\pi}{\cal F}(\Omega_2) &\approx&  \int_{ \Omega_2} 
  |\psi|^2|{\vec{\nabla}}\chi - {\vec A}|^2 
 + \frac{(1 - |\psi|^2)^2}{2}  \nonumber \\ 
 &\approx&  \frac{\delta}{2a} \left(\psi_0^2 \langle v_s^2  \rangle
 +( 1 - \psi_0^2)^2 \right) 
\label{edg1} 
\end{eqnarray}
where
\begin{equation}
\langle v_s^2  \rangle = 
\int \frac{d\theta}{2\pi}  |{\vec \nabla}  \chi - {\vec A}(R)|^2 
\label{avervel}
\end{equation}
is the superfluid velocity square averaged over the boundary of the disk.
 As before,  we have replaced the integral over $\Omega_2$ 
 by a line integral along the boundary of the sample ({\it i.e.} 
 the disk of radius $R$) multiplied by an effective  
 length $\delta.$  The function $\chi$ appearing in (\ref{edg1}) 
 is the phase of the order parameter, and the  vector  potential 
  is taken, as before, to  its value on the boundary 
 of the sample.  Optimizing 
 (\ref{edg1}) with respect to $\psi_0$ we find that: 
\begin{eqnarray} 
     \psi_0^2 &=&  1 - \frac{\langle v_s^2  \rangle}{2}
 \label{optpsi} \\ 
   \frac{1}{2\pi}{\cal F}(\Omega_2) &=& \frac{\delta}{2a}
\left(\langle v_s^2  \rangle-\frac{\langle v_s^2\rangle^2}{4} \right)
 \label{enomeg2} 
\end{eqnarray} 
for $\langle v_s^2  \rangle\le\sqrt{2}$.
 The phase function $\chi$ and the vector  potential near the 
 edge  of the disk are calculated  in Appendix B. 
 Using these  results, we obtain (for n=1): 
\begin{equation} 
{1 \over {2 \pi}} {\cal F}(\Omega_2) = \frac{\delta}{2a} \left(  a ( 1 -\phi_e )^2 - 
{a^3 \over 4} ( 1 -\phi_e )^4  \right) + f(x, a , \phi_e - 1) \delta 
\label{edg4} 
\end{equation} 
 The function $f(x, a, \phi_e - 1)$ determines 
 the dependance of the free energy on  the position $x$ of the vortex; 
 hence, it measures  the interaction 
 energy between the edge currents and 
  the  vortex as a function of its position. It  is given by 
\begin{equation} 
 f(x, a, \phi_e - 1) = \frac{ 2 a x^2}{  1 -  x^2} (\phi_e - 1)^2 
  \left(  1 -  a^2  \frac{(\phi_e - 1)^2 }{  1 -  x^2} 
       \right) 
 \label{barriere} 
 \end{equation} 
 From this expression, we observe  that the edge currents tend to 
  confine the vortex inside the system. 
 In figure 9 the surface energy as a function of the 
 position $x$ of the vortex is plotted. According to (\ref{optpsi}), only the 
 increasing part of the curve is physical. We nevertheless plot  the curve 
 defined by  (\ref{barriere})  in the whole range $ 0 \le x \le 1$ 
 in order to  emphasize the similarity between our result and 
 the well-known Bean-Livingston surface barrier 
 effect that was first derived using the London theory \cite{bean,degennes}.

 \section{Conclusion} 
 
   In this work, we have obtained analytical results for 
 the free energy and the magnetization 
  of a mesoscopic superconductor. We have used a known  exact solution 
 for the two dimensional Ginzburg-Landau equations in an infinite 
 plane, valid at the dual point, to  study  a finite system 
 with boundaries. With the help of numerical simulations, we have carried 
 out a perturbative calculation in the vicinity of the dual point. 
 This approach  enabled us to study thermodynamically stable states 
 but also metastable states (to obtain a surface energy barrier). This 
 model gives  theoretical insights into  the physical mechanisms involved 
 in the experimental results of \cite{geim1,geim2} and our 
 analytical results agree quantitatively with 
 experimental measurements.  In fact,  other related thermodynamic quantities 
 such as the surface tension measuring the thermodynamic stability  
of vortex states can also be computed along this way and could generalize to two-dimensional  
systems previous results obtained in one dimension \cite{dorsey}.

 More generally, we believe that a theoretical study in the vicinity 
 of the dual point provides  a lot  of information 
 about the  Ginzburg-Landau equations. 
 Although one usually relies on exact results derived from London's 
 equation, one should be aware of the fact that  these results agree 
 with numerical simulations of  Ginzburg-Landau equations only when 
 $\kappa$ is   large (typically $\kappa \ge 50).$ We  verified 
 that the behaviour we found in the vicinity of the dual point, 
 such as the scaling of the free energy, remains valid when 
  $\kappa$ ranges from 0.1  to 10 and this interval of values is indeed  
 relevant for  many conventional superconductors. 
 
 Our study can be extended in many directions. The scaling results 
 in the vicinity of $\kappa = 1/\sqrt{2}$  were derived from 
  numerical simulations: 
  a systematic perturbative expansion around the dual point 
 would put them on a more rigorous basis. Secondly, a linear stability 
 analysis  of the cylindrically symmetrical  solution \cite{bogo2} 
  should allow to understand 
 the fragmentation transition between a giant vortex and unit 
 vortices.  Since  the separatrix  $\Gamma$ exists even 
 for vortex configurations  breaking   cylindrical symmetry, our approach 
  can be used to analyze   hysteretic behaviour 
  of  metastable states, and to study  polygonal vortex 
 configurations found numerically in mesoscopic superconductors 
 \cite{deo,pala1}. 
 
\section{Acknowledgments} 
 
It is pleasure to thank G. Dunne for numerous discussions. 
K.M.  would like to express his gratitude to S. Mallick for his 
constant help during the preparation of this work and to thank 
A. Lemaitre for many interesting discussions. 
During the completion of this work, we received a preprint  
 by G.S. Lozano et al. \cite{lozano} which contains results similar to 
ours for the case of non interacting vortices.
It is a good opportunity to thank Gustavo Lozano for correspondence about 
his results.

This research was supported in 
part by 
the U.S.--Israel Binational Science Foundation (BSF), by the Minerva 
Center for Non-linear Physics of Complex Systems, by the Israel 
Science Foundation, by the Niedersachsen Ministry of Science 
(Germany) and by the Fund for Promotion of Research at the 
Technion.

   \section{Appendix A: The Ginzburg-Landau equations in a cylindrically 
 symmetric system} 
 
   For a  cylindrically  symmetric system, 
 we can use $ \psi =  f(r) e^{i n \theta}$ and  
$ { \vec  A} =  A(r) { \hat e_{\theta} }$  
 where $n$ is a non-negative integer which represents the number 
 of vortices  at the center of the system. We also define the 
 superfluid velocity $\vec{v}_s =v_s (r) { \hat e_{\theta} }$, where 
\begin{equation} 
v_s (r) =\left(\frac{n}{r}-A(r)\right) 
\label{vs} 
\end{equation} 
In this case the Ginzburg-Landau equations are: 
\begin{eqnarray} 
  \frac{ d^2 f}{dr^2}  + \frac{1}{r}  \frac{ d f}{dr} 
 - v_s^2 f &=& - 2 \kappa^2 f ( 1 - f^2)  \\ 
 \frac{ d }{dr}\left(\frac{1}{r} \frac{ d }{dr} (r 
 v_s)\right)&=&2v_s f^2 
\end{eqnarray} 
It is convenient to define the quantity $p(r) = r v_s(r)$. 
The magnetic field $\vec{B} = B(r)  \hat{e}_z$ is given in terms of $p(r)$ 
by 
\begin{equation} 
B(r)= -\frac{1}{r}\frac{dp}{dr} 
\label{magnfield} 
\end{equation} 
We obtain finally two coupled ordinary differential equations 
 \begin{eqnarray} 
   f'' &=& - 2 \kappa^2 f ( 1 - f^2)+p^2 f^2/r^2-f'/r  \\ 
   p'' &=& 2 p f^2+p'/r 
\label{appgl2} 
\end{eqnarray} 
with the following boundary conditions at   $ r = a^{-1}$ for $n\neq 0$: 
\begin{equation} 
\begin{array}{ll} 
 f(0) =  0\;\;\;\;\;\;\;\; & f'(a^{-1}) = 0  \\ 
 p(0) = n  & p(a^{-1}) = n-\phi_e 
\end{array} 
\end{equation} 
for a disk and 
\begin{equation} 
\begin{array}{ll} 
 f(0) =  0\;\;\;\;\;\;\;\; & f'(a^{-1}) = 0  \\ 
 p(0) = n  & p'(a^{-1}) = -2 a \phi_e 
\end{array} 
\end{equation} 
for a cylinder. These are the equations we have  solved  numerically using the 
relaxation method \cite{numrec}. From the analysis of the equations (\ref{appgl2}) 
we  deduce the following behaviour in the vicinity of the center of the disk: 
$$  f \sim r^n  \,\,\,\,  \hbox{  and }   \,\,\,\, p \sim r^2 
     \hbox{  when  }   r \to 0 $$ 
 
The free energy (\ref{energ}) is then given in terms of the solution of (\ref{appgl2}) by 
\begin{equation} 
 \frac{\cal F}{2\pi}=\int_{0}^{1/a} rdr\,\left(\frac{B^2}{2}+\kappa^2\left(1-f^4\right) \right)  
\end{equation}

   \section{Appendix B: phase and vector potential 
of an off centered  configuration  with one vortex} 
 
In this appendix we measure 
the distances in units of $R$, so the disk has unit radius.
Suppose  that the vortex is located at a distance 
$ x $ from the center of the disk $(0 \le  x < 1)$. 
The phase $\chi(\rho, \theta)$ of the order parameter satisfies 
$\Delta\chi = 0$ everywhere  on  the disk except  on the vortex 
with boundary condition ${\hat {\bf n}} . \vec{\nabla}\chi  = 0 $
 
Using  the image method, the phase $\chi(\rho,\theta)$   at a point 
located at a distance $\rho$ from the center of the disk 
 (with $ 0 \le  \rho  \le  1$) is  given by \cite{fetter}: 
\begin{equation} 
   \chi(\rho,\theta) = 
  \mbox{Im}\ln \left(\frac{ \rho \exp(i\theta) -x} 
{ \rho \exp(i\theta) - x^{-1}}\right) 
\label{phase} 
\end{equation} 
 where Im  denotes the imaginary part part of a complex-valued function. 
 Or equivalently: 
 \begin{equation} 
  \tan \chi(\rho,\theta) = \frac{ 1 - x^2}{ 1 + x^2} \,\, 
 \frac{  \sin\theta} {\cos\theta - \frac{\rho+\rho^{-1}}{x + x^{-1}}} 
 \label{phase1} 
  \end{equation} 
    On the boundary of the disk, $\rho=1$, and one finds that 
 \begin{equation} 
   \frac{\partial\chi}{\partial\theta} =   \frac{ 1 - x^2}{ 1 + x^2} 
  \frac{1}{ 1 - \frac{2x}{1 + x^2} \cos\theta},\;\;\;\;\; 
   \frac{\partial\chi}{\partial\rho} =   0 
  \end{equation} 
therefore
\begin{equation}
\int \frac{d\theta}{2\pi} |{\vec \nabla}  \chi(1, \theta) |^2 
 =  a^2   \frac{ 1 +  x^2}{ 1 -  x^2} 
\end{equation}

 The vector-potential ${\vec A}(R)$ at the boundary of the sample 
  is   a function of the polar angle $\theta$ since 
 the vortex is not  at the center of the disk. 
 We   determine ${\vec A}(R)$   from the following conditions: 
$${\vec \nabla}.{\vec A}= 0,  \,\,\,\,\, \oint_{\partial\Omega} 
 {\vec A}(R).d{\vec l} = \phi_e 
$$
and on the boundary  $\vec{A}(R). \hat{\bf n} = 0$. 
The following choice,  
 \begin{equation} 
  {\vec A}(R) = \phi_e  {\vec \nabla} \chi 
\end{equation} 
valid near boundary of the system, 
 satisfies these  requirements.

\begin{figure} 
\centerline{\psfig{figure=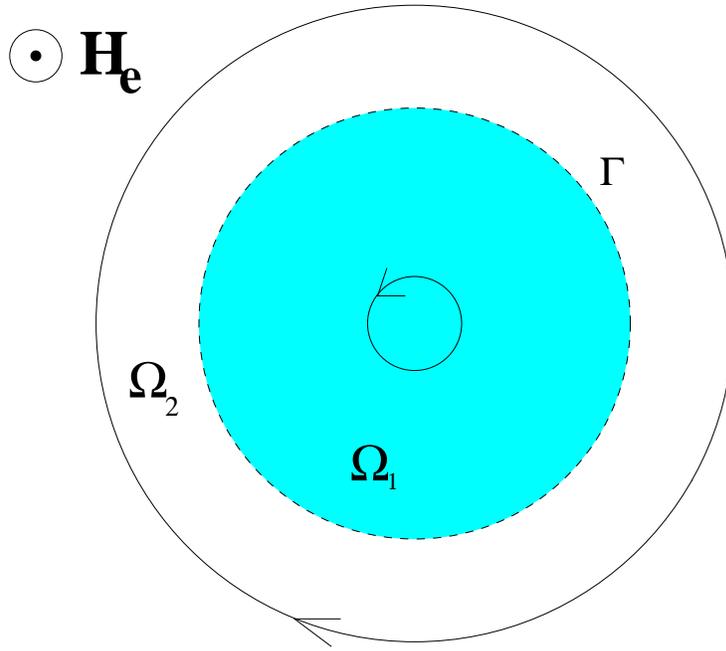,width=8.6cm,angle=270}}
\hspace{10pt}
\caption{The sample cross-section is divided into two subdomains 
 by the circle $\Gamma$. The arrows indicate the direction of the current.  }
\label{domains}  
\end{figure} 
\begin{figure} 
\centerline{\psfig{figure= 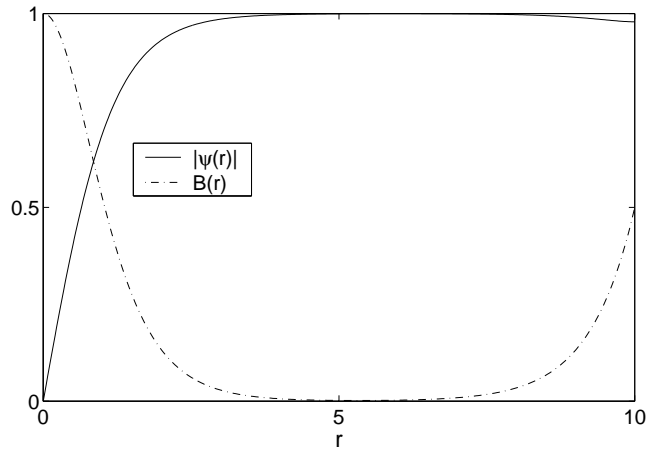,width=8.6cm,angle=0}} 
\hspace{10pt}
\caption{Behaviour  of the order parameter 
 and the magnetic field  at the dual point  for 
 a  cylinder of radius  10  $\lambda \sqrt 2$ 
  containing one vortex. } 
\label{dima1} 
\end{figure} 
 \begin{figure}
\centerline{\psfig{figure=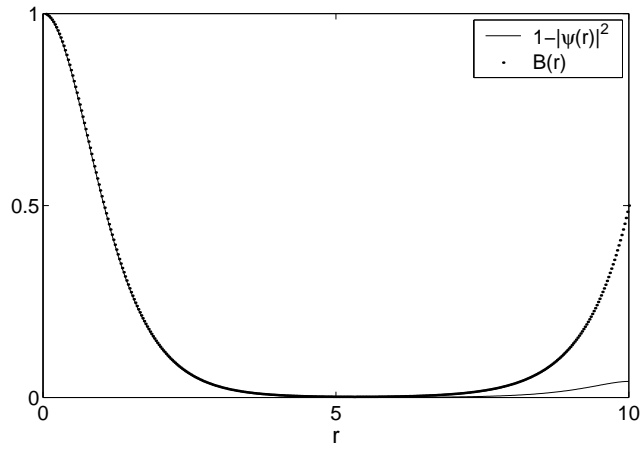 ,height=6cm,angle=0}} 
\hspace{10pt} 
\caption{ Comparison between $B$ and $1 - |\psi|^2$ in a cylinder 
    of radius  10  $\lambda \sqrt 2$ 
   containing one vortex. } 
\label{dima2} 
\end{figure} 
\begin{figure} 
\centerline{\psfig{figure=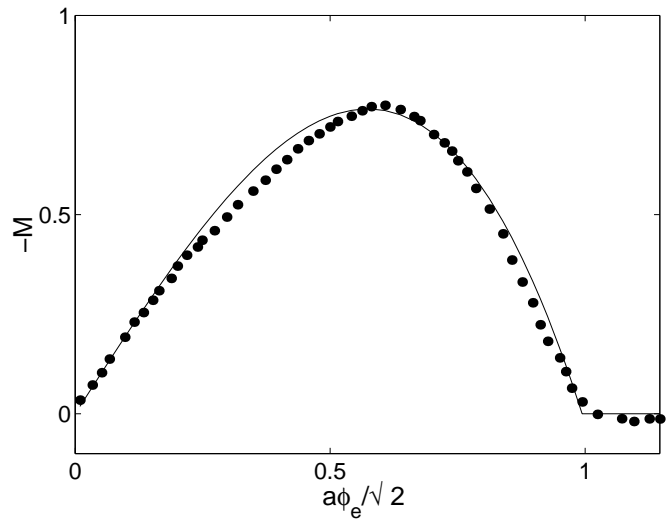,height=7cm,angle=0}} 
\hspace{10pt} 
\caption{Magnetization of a fractional fluxoid disk. Comparison 
 between the experimental measurements [8] \ (for $R = 0.31 \mu m$)  
and the theoretical curve taken from the expression (\ref{mag0}). }
\label{cubic} 
\end{figure} 
\begin{figure} 
\centerline{\psfig{figure=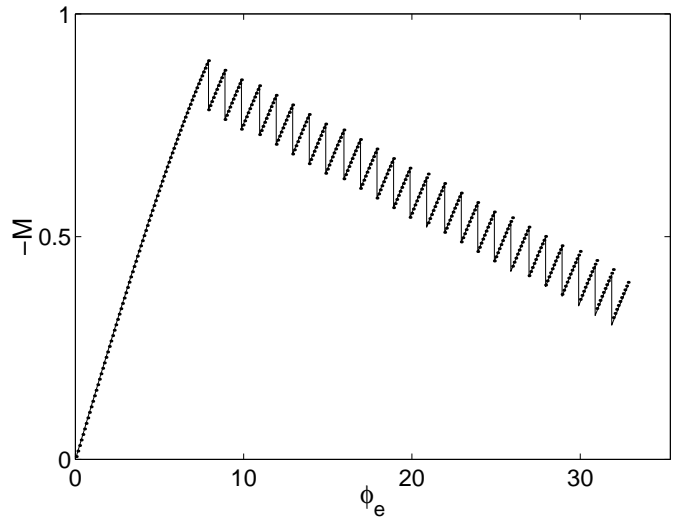,height=7cm,angle=0}} 
\hspace{10pt}
\caption{Behaviour of the magnetization of 
 a disk with radius 10 $\lambda \sqrt 2$, at the dual point. Dots represent the numerical solution  
and the solid curve the expression (\ref{magthermo}) together with (\ref{enlibre}). The only  
free parameter $\delta$ has been taken to $\delta = 0.76 \lambda$. } 
\label{ptdual} 
\end{figure} 
\begin{figure} 
\centerline{\psfig{figure=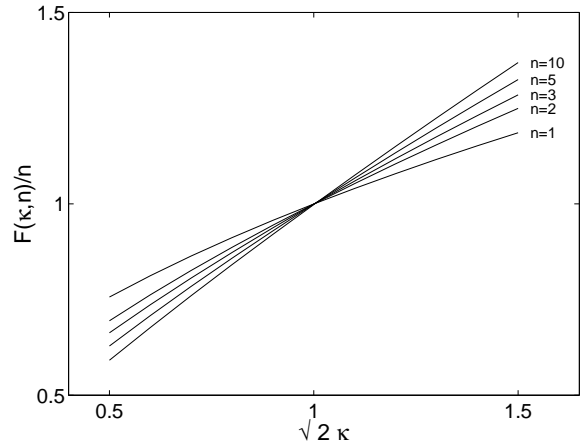,height=6cm,angle=0}}
\hspace{10pt} 
\caption{Behaviour of the free energy per vortex
$F/n={\cal F}/2\pi n$ as a function of 
  $\sqrt 2 \kappa$ 
 for different values of $n$, the number  of vortices. 
 At the self-dual point $\sqrt 2 \kappa =1$, 
 the energy ${\cal F}(n) = n {\cal F}(1)$ so that the 
 interaction energy between the vortices vanishes 
identically.  } 
\label{jacobs} 
\end{figure} 
\begin{figure} 
\centerline{\psfig{figure= 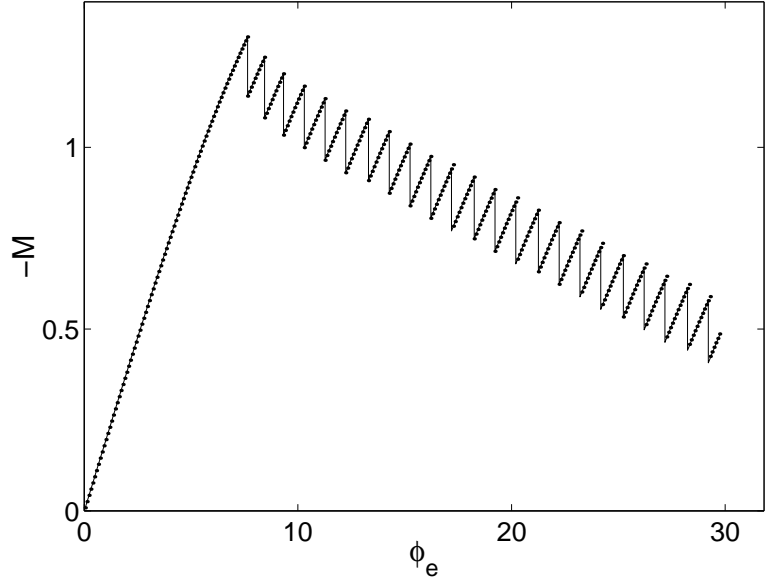,height=8cm,angle=0}} 
\hspace{10pt}
\caption{Magnetization curve of   a disk of radius 10 $\lambda \sqrt 2$, 
   as a function of the applied field 
 for $\kappa {\sqrt 2} = 0.9$. Dots represent the numerical solution  
and the solid curve the expression (\ref{magthermo}) together with (\ref{pertbord}, \ref{fscaling}). The only  
free parameter $\delta$ has been taken to $\delta = 0.76 \lambda$.}
\label{intercafig}  
\end{figure} 
\begin{figure} 
\centerline{\psfig{figure=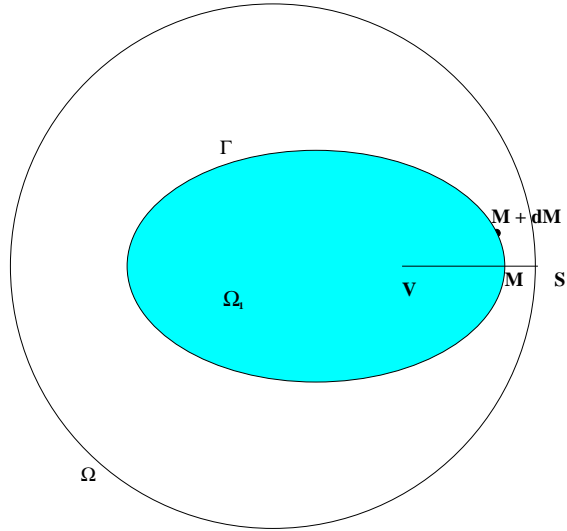,height=7cm,angle=0}} 
\hspace{10pt}
\caption{The separation of a system  without cylindrical symmetry 
 in two subdomains by a  curve  $\Gamma$.  } 
\label{courbegamma} 
\end{figure} 
\begin{figure} 
\centerline{\psfig{figure=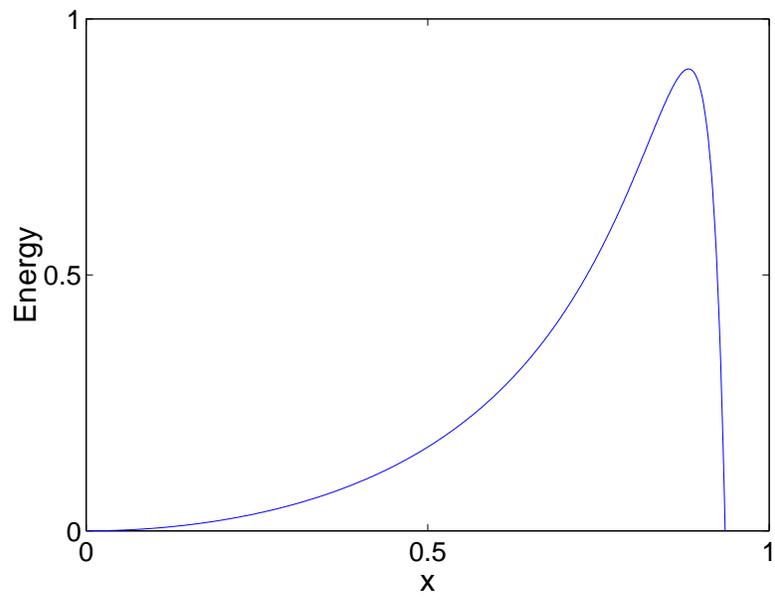,height=8cm,angle=0}} 
\hspace{10pt}
\caption{Confining energy of a vortex inside a disk due to  
 edge currents.  }\label{beanfig}  
\end{figure} 

\newpage
\begin{table}
  \centering 
  \begin{tabular}{||c|c||c|c||c|c|} 
  \hline 
 &                &         &                &      &    \\ 
 n   &   $\alpha (n)$ &     n   &  $\alpha(n)$   &  n   &  $\alpha(n)$\\ 
     &                &         &                &      &    \\ 
\hline 
1   &    0.417       &    11   &   0.785        &  21  &    0.841  \\ 
2   &    0.544       &    12   &   0.794        &  22  &    0.845  \\ 
3   &    0.613       &    13   &   0.802        &  23  &    0.847  \\ 
4   &    0.658       &    14   &   0.809        &  24  &    0.850  \\ 
5   &    0.690       &    15   &   0.815        &  25  &    0.853  \\ 
6   &    0.715       &    16   &   0.821        &  26  &    0.855  \\ 
7   &    0.734       &    17   &   0.826        &  27  &    0.857  \\ 
8   &    0.750       &    18   &   0.830        &  28  &    0.859  \\ 
9   &    0.764       &    19   &   0.834        &  29  &    0.860  \\ 
10  &    0.775       &    20   &   0.838        &  30  &    0.862  \\ 
\hline 
\end{tabular}
\hspace{10pt} 
\caption{The numerical values of the function $\alpha (n)$ for  $n$ ranging from $1$ to $30$.} 
 \label{alpha}
\end{table}


\begin{thebibliography}{article} 
\bibitem{geim1}  A.K. Geim, I.V. Grigorieva, 
 S.V. Dubonos, J.G.S. Lok, J.C. Maan, A.E. Filippov and F.M. Peeters, 
 Nature, {\bf 390}, 259 (1997) 
\bibitem{dalibard} K.W. Madison, F. Chevy, W. Wohlleben and J. Dalibard, 
 to appear in Phys.Rev.Lett  cond-mat/9912015 
\bibitem{geim0}  A.K. Geim,  S.V. Dubonos, J.G.S. Lok,  I.V. Grigorieva, 
 J.C. Maan, L. Theil Hansen and P.E. Lindelof, Appl.Phys.Lett. {\bf 71}, 
 2379 (1997). 
\bibitem{geim2}   A.K. Geim, S.V. Dubonos, J.G.S. Lok, 
 M. Henini and  J.C. Maan,  Nature, {\bf 396}, 144  (1998) 
\bibitem{moschalk} V.V. Moshchalkov, L. Gielen, C. Strunk, R. Jonckheere, 
 X. Qiu, C. Van Haesendonck and Y. Bruynseraede, 
 Nature, {\bf 373}, 319  (1995) 
\bibitem{buisson} O. Buisson, P. Gandit, R. Rammal, Y.Y. Wang and 
 B. Pannetier, Phys.Lett.A, {\bf 150}, 36 (1990) 
\bibitem{peetreview}  P. Singha Deo, F.M. Peeters  and   V.A. Schweigert, 
 submitted to Superlattices   and  microstructures (1999) 
 cond-mat/9812193 
\bibitem{deoprl1} P. Singha Deo, V.A. Schweigert, F.M. Peeters and A.K. 
 Geim,  Phys.Rev.Lett. {\bf 79}, 4653 (1997) 
\bibitem{deo}  P. Singha Deo, V.A. Schweigert and F.M. Peeters 
   Phys. Rev. {\bf B 59}, 6039  (1999) 
\bibitem{pala1} J.J. Palacios, Phys. Rev. {\bf B 58}, R5948  (1998) 
\bibitem{pala2}  J.J. Palacios, cond-mat/9908341 
\bibitem{schweig}  V.A. Schweigert,  P. Singha Deo  and  F.M. Peeters, 
  Phys.Rev.Lett. {\bf 81}, 2783  (1998) 
\bibitem{degennes}  P.G. de Gennes, {\em Superconductivity of metals and alloys} 
 Addison-Wesley (1989) 
\bibitem{little} W. A.  Little and R.D. Parks, 
 Phys.Rev.Lett. {\bf 9}, 9  (1962) 
\bibitem{groff} R. P. Groff and R.D. Parks, Phys.Rev {\bf 176}, 567  (1968) 
\bibitem{zwerger}  R. Benoist and W. Zwerger, Z.Phys. {\bf B 103}, 377 (1997) 
\bibitem{bobel}   G. B{\"o}bel,  Nuovo Cimento  {\bf 38}, 
 6320 (1965) 
\bibitem{shapoval}  E.A. Shapoval, JETP Letters {\bf 69}, 577 (1999) 
\bibitem{pearl}  J. Pearl,  Appl. Phys. Lett. {\bf 5}, 65 (1964) 
\bibitem{fetter}  A.L. Fetter, Phys. Rev. {\bf B 22}, 1200 (1980) 
\bibitem{bean} C.P. Bean and J.D.   Livingston, Phys.Rev.Lett. 
{\bf 12}, 14  (1964) 
\bibitem{russes} A.S. Krasilnikov, L.G. Mamsurova, N.G. Trusevich, 
 L.G. Shcherbakova and K.K Pukhov, Supercond.Sci.Technol {\bf 8}, 1 (1995) 
 \bibitem{bresiliens}  P.A. Venegas and E. Sardella, Phys. Rev. {\bf B 58}, 
  5789 (1998) 
 \bibitem{sarma} D. Saint-James, E.J. Thomas and G. Sarma, {\em Type II 
 Superconductivity}  Pergamon Press  (1969) 
\bibitem{bogo} E.B. Bogomol'nyi, Sov.J.Nucl.Phys. {\bf 24}, 449 (1977) 
\bibitem{peet2}  V.A. Schweigert and  F.M. Peeters,  Phys.Rev.Lett. 
 {\bf 83}, 2409 (1999);  V.A. Schweigert and  F.M. Peeters, cond-mat/9910110 
\bibitem{harden} J.L. Harden and V. Arp, Cryogenics {\bf 3}, 105 (1963) 
\bibitem{akmal}  E. Akkermans and  K. Mallick 
  {\it  J. Phys. A.} {\bf 32},  7133 (1999) 
\bibitem{rebbi} L. Jacobs and C. Rebbi, Phys.Rev. {\bf B 19}, 4486 (1979) 
\bibitem{london} F. London, {\em Superfluids, Vol. 1: Macroscopic 
 theory of superconductivity} Dover (1960) 
\bibitem{dunne} G. Dunne, private communication. 
\bibitem{houches} E. Akkermans  and K. Mallick, proceedings 
   of  Les Houches Summer School {\em Topological Aspects of 
 Low dimensional systems}  (1999) cond-mat/9907441 
\bibitem{schweig2} V.A. Schweigert and  F.M. Peeters, Phys.Rev {\bf B 57}, 
 13817  (1998) 
\bibitem{dima}  E. Akkermans, D. Gangardt  and K. Mallick, 
 in preparation 
\bibitem{bogo2}  E.B. Bogomol'nyi and A.I.  Vainshtein, Sov.J.Nucl.Phys 
 {\bf 23}, 588 (1976) 
\bibitem{dorsey} A.T. Dorsey, Ann.Phys, {\bf 233}, 248 (1994) 
\bibitem{lozano} G.S. Lozano, M.V. Manias and E.F. Moreno, cond-mat/0005199
\bibitem{numrec}  W.H. Press, B.P. Flannery, S.A. Teukolsky and W.T. Vetterling, 
{\em Numerical Recipes: The Art of Scientific Computing } 
 Cambridge University Press, (1992) 
\end{thebibliography}
\end{document}